\documentclass[11pt]{article}
\usepackage{enumerate}
\usepackage{amsmath}
\usepackage{amssymb}
\usepackage{bm}
\usepackage{ascmac}
\usepackage{amsthm}
\usepackage[dvipdfmx]{graphicx,color}
\usepackage{braket}
\usepackage{tikz}
\usetikzlibrary{calc}

\usepackage{a4wide}
\usepackage{cases}
\usepackage{mathrsfs}
\usepackage{authblk}

{\theoremstyle{plain}%
  \newtheorem{thm}{\bf Theorem}[section]%
  \newtheorem{lem}[thm]{\bf Lemma}%
}
{\theoremstyle{remark}
  
}

\newcommand{\dC}{{\mathbb{C}}}%
\newcommand{\dN}{{\mathbb{N}}}%
\title{Quantum search of matching on signed graphs}
\author[1]{Etsuo SEGAWA\footnote{E-mail: segawa-etsuo-tb@ynu.ac.jp}}
\author[2]{Yusuke YOSHIE\footnote{Corresponding author, E-mail: yoshie@math.gakushuin.ac.jp}}
\affil[1]{Graduate School of Environment and Information Sciences, Yokohama National University, \newline 79-1, Hodogaya, Yokohama 240-8501, Japan}
\affil[2]{Department of Mathematics Faculty of Sciences Gakushuin University,\newline 1-5-1, Mejiro, Toshima, Tokyo 171-8588, Japan}
\date{}
\begin {document}
\maketitle{}
\begin{abstract}
We construct a quantum searching model of a signed edge driven by a quantum walk. The time evolution operator of this quantum walk provides a weighted adjacency matrix induced by the assignment of sign to each edge. This sign can be regarded as so called the edge coloring. Then as an application, under an arbitrary edge coloring which gives a matching on a complete graph, we consider a quantum search of a colored edge from the edge set of a complete graph. We show that this quantum walk finds a colored edge within the time complexity of $O(n^{\frac{2-\alpha}{2}})$ with probability $1-o(1)$ while the corresponding random walk on the line graph finds them within the time complexity of $O(n^{2-\alpha})$ if we set the number of the edges of the matching by $O(n^{\alpha})$ for $0 \le \alpha \le 1$.

 \end{abstract}
\section{Introduction}
A quantum walk is introduced as a quantum analogue of a classical random walk \cite{Gudder} and the fundamental idea can be seen in \cite{Feynman}. As Y. Aharonov \cite{Aharonov} {\it et al.} formulated quantum random walks as an antecedent of the current quantum walk and Meyer \cite{Meyer} introduced it as a quantum cell automaton, quantum walks have been in the limelight for the last two decades.  One of the reason why quantum walks have been intensively studied for a long time is that the quantum walks often exhibit a specific characteristic which cannot be seen in classical random walks in both of infinite, and finite graphs. As is seen in \cite{Amb} and \cite{Nayak}, by a combinatorial and the 
Schr\"odinger's approach, the asymptotic probability distribution of a quantum walk in an infinite graph was analyzed, which shows a definitely different behavior from that of a classical random walk. Thereby, Konno \cite{Konno02}, \cite{Konno05} obtained the weak limit distribution of a quantum walk which corresponds to the central limit theorem of a classical random walk. Then the time scaling for the quantum walk to obtain the weak limit is linear in time while  the time scale to obtain the limit for the random walk is the square root of time. It implies that a quantum walker linearly spreads. The property is called the {\it linear spreading}. Moreover, {\it localization} \cite{Inui04} is also well-studied characteristic of quantum walks on infinite graphs seen in some classes, e.g., one-dimensional lattice \cite{Konno10}, the half line \cite{Konno11} and so forth. 

In finite graphs, D. Aharonov \cite{DAharonov} {\it et al.} formulated quantum mixing, filling and dispersion time and showed that these are quadratically faster than classical ones. Constructing an efficient system to find targets in a graph has been one of main research topic of quantum walks. In other words, quantum walks enable us to propose an algorithm finding marked elements in a graph efficiently. For example, the original Grover's search algorithm \cite{Grover} can be regarded as a quantum search on  the complete graph $K_{N}$ with self loops driven  by a quantum walk and  it finds marked items within the time complexity of $O(\sqrt{N})$ while that of a classical searching algorithm  is $O(N)$. In that case, a quantum walk gives quadratic speed-up to the searching algorithm. Besides this, quantum searchings are considered in some classes of graphs, e.g., a finite $d$-dimensional grid \cite{Amb3}, hypercubes \cite{Shenvi}, triangular lattices \cite{Abal}, and Johnson graphs \cite{Amb2}. Furthermore, Matsue \cite{Matsue18} {\it et al.} proposed a quantum search on simplicial complexes as an extension of graphs. It was shown that these models also provide speed-up to find marked elements. A principal idea to construct a quantum search algorithm is driving a quantum walk with a perturbation. In other words, we observe marked elements with sufficiently high probability by proper times applications of a perturbed quantum walk.   Details are seen in, e.g., \cite{Aaronson}, \cite{Magniez}, and \cite{Sze04}.

In this paper, we give a perturbation to the edge set of a complete graph by a {\it sign}. This sign is a map from the set of edges to $\{ \pm 1 \}$ which corresponds to an edge coloring. We call such a graph a {\it signed graph}. Harary \cite{Harary} introduced signed graphs as a model for a social network.  Brown \cite{Brown} {\it et al.} proposed perfect state transfer, which is a specific property of a continuos-time quantum walk in a signed graph and studied how negatively signed edges effect to the perfect state transfer. Here, we call the negatively signed edges the {\it marked edges} and regard them as targets of our searching model. Through the research, we aim to see how the existence of negatively signed edges affects to our searching model. The idea of this paper is constructing a perturbed quantum operator $U_{\sigma}$ and finding marked edges in a signed complete graph as fast as possible by driving $U_{\sigma}$. Here, $U_{\sigma}$ is constructed by a {\it weighted adjacency matrix} $T_{\sigma}$ of a complete graph induced by the edge coloring.  See Section 2 for the detailed definition. The spectral analysis on $T_{\sigma}$ is studied by Akbari \cite{Akbari} {\it et al.}  if $\sigma$ accomplishes a {\it matching} which is a set of disconnected edges. Then we  treat the case where the set of marked edges becomes a  matching. We call a matching having $t$ edges a $t$-matching. If the matching covers all the vertices, it is called a {\it perfect matching}. More precisely, we search a signed $t$-matching on the complete graph on $n+1$ vertices. Note that the number of edges of the perfect matching is $\left[ \frac{n+1}{2} \right]$. Then we set the number of edges in the matching by $t=O(n^{\alpha})$ for $0 \le \alpha \le 1$. In addition, we will show that the probability finding marked edges is sufficiently high after driving the perturbed quantum walk within the time complexity of $O(n^{\frac{2-\alpha}{2}})$ while a classical one requires the time complexity of $O(n^{2-\alpha})$. Thus, the quantum algorithm proposed in this paper gives quadratic speed-up.  The following statement is our main result in this paper. 
\begin{thm}
For a sufficiently large $n$, quantum search driven by $U_{\sigma}$ in a signed complete graph on $n+1$ vertices enables us to find a marked $t$-matching with the time complexity of $O(n^{\frac{2-\alpha}{2}})$, where $t=O(n^{\alpha})$ for $0 \le \alpha \le 1$.  
\label{search on Kn+1}
\end{thm}

This paper is organized as follows. In Section 2, we define a {\it sign} on graphs and give definition of {\it marked} edges. In addition, we construct a time evolution operator of the perturbed quantum walk by the sign. In Section 3, we estimate searching time finding marked edges by a classical searching algorithm by spectral analysis. In Section 4, we consider quantum searching on a signed complete graph and prove our main result, that is, we show that the searching time on the graph given by our quantum walk is quadratically faster than that of a classical searching algorithm. Section 5 is devoted to summarizing our results and make discussion for our future direction. 

\section{Preliminaries}
\subsection{Graph and sign}
Let $G=(V,E)$ be a connected and simple graph. For $e=uv \in E$, the vertices $u$ and $v$ are called the {\it endpoints} of $e$. If two edges $e$ and $f$ with $e \ne f$ share a vertex as their endpoints, that is, $e=uv$ and $f=wx$ satisfy $v=w$, we write $e \approx f$. In addition, if $u \in V$ is an endpoint of $e \in E$, we denote by $u \in e$. The oriented edge from $u \in V$ to $v \in V$ is called an {\it arc} and denoted by $(u, v)$.  Define $\mathcal{A}=\{ (u,v), (v,u) \mid uv \in E\}$ which is the set of the symmetric arcs of $G$. For $a \in \mathcal{A}$, $t(a)$ and $o(a)$ denote the terminus and origin of $a$, respectively. In addition, $a^{-1}$ denotes the inverse arc of $a$. 
Moreover, we write the adjacency matrix of a graph $G$ as $A(G)$.  Furthermore, for $r \in \dN$, the zero-vector and the all-one vector in $\dC^{r}$ is denoted by ${\bf j}_{r}$ and ${\bf 0}_{r}$, respectively. Throughout the paper, $\mathrm{Spec}(X)$ denotes the set of eigenvalues of a matrix $X$.  

Before defining a sign, let us introduce some classes of graphs. For $r \in \dN$, the {\it complete graph} on $r$ vertices, denoted by $K_{r}$, is a graph in which every pair of distinct two vertices are adjacent.  For $r \in \dN$, the {\it cocktail party graph}, denoted by $\mathrm{CP}(r)$, is the graph in which the set of the vertices is decomposed into $r$ distinct subsets $V_{1}, V_{2}, \dots, V_{r}$ with $
|V_{i}|=2$ for $1 \le i \le r$ and every pair of two vertices are connected unless these are belong to the same partite set. In other words, the adjacency matrix of $K_{r}$ and $\mathrm{CP}(r)$ is given by
\[ 
A(K_{r})=\left(
\begin{array}{ccccc}
0 & 1 & 1 & \dots &1\\
1 & 0 & 1 & \dots &1\\
1 & 1 & 0 & \dots &1\\
\vdots & & \ddots & &\vdots \\
 1 & 1 & \dots & 1 & 0
\end{array}
\right),
\]
and
\[ 
A(\mathrm{CP}(r))=
\left(
\begin{array}{ccccccc}
0 & 0 & 1 & 1& \dots &1& 1\\
0 & 0 & 1 & 1 & \dots &1 & 1\\
1 & 1 & 0 & 0 & \dots &1 & 1\\
1 & 1 & 0 & 0 & \dots &1 & 1\\
\vdots & & \ddots & &\vdots \\
1 & 1 &1 & 1 & \dots & 0 & 0\\
1 & 1 &1 & 1 & \dots & 0 & 0
\end{array}
\right)=A(K_{r}) \otimes \left(
\begin{array}{cc}
1 & 1\\
1 & 1
\end{array}
\right),
\]
respectively. We call the set of disconnected edges a {\it matching}. In other words, a matching is the set of edges in which every pair of two edges $(e,f)$ satisfies $e \not \approx f$. If the number of edges in a matching is $t$, it is called a $t$-matching. 

Let $\sigma: \mathcal{A} \to \{ \pm 1 \}$ be a map satisfying that $\sigma(a^{-1})=1$ whenever $\sigma(a)=-1$ for $a \in \mathcal{A}$. Define $\mathcal{M}=\{a \in \mathcal{A} \mid \sigma(a)=-1\}$ and $\mathcal{M}^{-1}=\{a \in \mathcal{A} \mid a^{-1} \in \mathcal{M}\}$.  From the above $\sigma$, we give a {\it sign} $\tau: E \to \{ \pm 1\}$ by
\[ 
\tau(uv)=
\begin{cases}
1, & \sigma((u,v))\cdot \sigma((v,u))=1,\\
-1, & \sigma((u,v))\cdot \sigma((v,u))=-1.
\end{cases}
\]  
In other words, $\tau(uv)=\sigma((u,v))\cdot \sigma((v,u))$. If $e \in E$ satisfies $\tau(e)=-1$, we call it a {\it marked edge}. Define $M=\{ e \in E \mid \tau(e)=-1\}$ which is the set of marked edges and  $\partial{M}=\{t(a), o(a) \in V \mid a \in \mathcal{M} \}$ which is the set of the endpoints of the marked edges. 

\subsection{Time evolution operator}
In this part, we construct a time evolution operator of the perturbed quantum walk from the above $\sigma$. First, let us define an operator $S$ on $\dC^{\mathcal{A}}$ by
\[
S_{a,b}=
\begin{cases}
1, & a=b^{-1},\\
0, & \text{otherwise}.
\end{cases}
\]
Note that $S^{2}=I_{\mathcal{A}}$. In addition, we give a boundary operator $d_{\sigma}: \dC^{\mathcal{A}} \to \dC^{V}$ by
\[
(d_{\sigma})_{v,a}=
\begin{cases}
w(a), & t(a)=v,\\
0, & \text{otherwise},
\end{cases}
\]
where
\begin{equation*}
 w(a)=
\begin{cases}
-\frac{1}{\sqrt{\deg(t(a))}}, & a \in \mathcal{M},\\
\frac{1}{\sqrt{\deg(t(a))}}, & \text{otherwise}.
\end{cases}
\end{equation*}
It follows immediately that 
\[
(d^{*}_{\sigma})_{a,v}=
\begin{cases}
w(a), & t(a)=v,\\
0, & \text{otherwise}.
\end{cases}
\]
Then it is easily checked that $d_{\sigma}d^{*}_{\sigma}=I_{V}$ and $d^{*}_{\sigma}d_{\sigma}$ is a projection operator. We define the time evolution operator of the quantum walk by
\[ U_{\sigma}:=S(2d^{*}_{\sigma}d_{\sigma}-I_{\mathcal{A}}).\]
Remark that $U_{\sigma}$ is a unitary operator on $\dC^{\mathcal{A}}$ because of $S^{2}=I_{\mathcal{A}}$ and $d_{\sigma}d^{*}_{\sigma}=I_{V}$. For an initial state $\varphi_{0} \in \dC^{\mathcal{A}}$, the time evolution of the quantum walk is given  by $\varphi_{t}=U^{t}_{\sigma}\varphi_{0}$ and the finding probability on an {\it edge} $uv$ at time $t$ is given by
\[ |\varphi_{t}((u,v))|^{2}+|\varphi_{t}((v,u))|^{2}. \]
Define $T_{\sigma}=d_{\sigma}Sd^{*}_{\sigma}$. It is checked that $T_{\sigma}$ is an operator on $\dC^{V}$ whose entry is
\begin{equation}
(T_{\sigma})_{u,v}=
\begin{cases}
\tau(uv)\cdot\frac{1}{\sqrt{\deg{u}\deg{v}}} & u \sim v,\\
0 & \text{otherwise}.
\end{cases}
\label{def of T}
\end{equation}
Here, it is known that a part of the spectrum of $U_{\sigma}$ is expressed in terms of that of $T_{\sigma}$. 
\begin{thm}[Higuchi-Konno-Sato-Segawa \cite{HKSS14}]
Let $U_{\sigma}$ and $T_{\sigma}$ be defined as in the above. Then it holds that 
\[ \{ e^{\pm i\theta_{\lambda}} \mid \lambda \in \mathrm{Spec}(T_{\sigma})\backslash\{ \pm 1\} \} \subset \mathrm{Spec}(U_{\sigma}),\]
where $\theta_{\lambda}=\cos^{-1}{\lambda}$.  In addition,  the normalized eigenvector of $U_{\sigma}$ associated to $e^{\pm i \theta_{\lambda}}$ is given by 
\begin{equation}
\varphi_{\pm\lambda}=\frac{1}{\sqrt{2}|\sin{\theta_{\lambda}}|}(d^{*}_{\sigma}f-e^{\pm i\theta_{\lambda}}Sd^{*}_{\sigma}f),  \label{eig func}
\end{equation}
where $f$ is the normalized eigenvector of $T_{\sigma}$ associated to $\lambda$. 
That is,
\begin{equation*}
\varphi_{\pm\lambda}(a)=\frac{1}{\sqrt{2}|\sin{\theta_{\lambda}}|}\left( w(a)f(t(a))-e^{\pm i\theta_{\lambda}}w(a^{-1})f(o(a))\right).
\end{equation*} 
\end{thm}

Here, we treat the case where $G$ is the complete graph on $n+1$ vertices and $M$ becomes a $t$-matching with $t=O(n^{\alpha})$ for $0 \le \alpha \le 1$.  We set $c>0$ to be the coefficient of $n^{\alpha}$ of $t$, that is, $t=[cn^{\alpha}]$. Note that $c$ is small enough to satisfy $cn^{\alpha} \le \frac{n}{2}$. 

\section{Classical search algorithm}
In this part, we estimate a classical searching time finding an edge in $M$ on a signed complete graph $G=K_{n+1}$.  
To this end, we introduce a simple random walk on edges of $G$ by the following operator $P$ on $\dC^E$: 
\[ P_{e,f}=
\begin{cases}
\frac{1}{2(n-1)}, & e \approx f,\\
0, & \text{otherwise.}
\end{cases}
\]
In other words, it holds that 
\[ P=\frac{1}{2(n-1)}A(L(G)),\]
where $L(G)$ is so called the {\it line graph} of $G$ whose definition is given by 
\begin{align*}
V(L(G))&=E(G), \\
E(L(G))&=\{ ef \mid \text{$e \approx f$ and $e \ne f$ in $G$} \}.
\end{align*}
Let us denote by $P_{M}$ the $(|E|-t) \times (|E|-t)$-matrix obtained by removing all the rows and columns of $P$ corresponding to the edges in $M$, that is, 
\[P_{M}=\frac{1}{2(n-1)}A(L(G_{M})), \] 
where $G_{M}$ is the graph obtained by removing all the edges in $M$. According to \cite{Broder}, the classical searching time finding edges in $M$, which is the expectation of the first hitting time to the matching is $O\left(\frac{1}{1-\mu_{m}}\right)$, where $\mu_{m}$ is the maximum eigenvalue of $P_{M}$. Thus, we will evaluate the maximum eigenvalue of $P_{M}$ to estimate the classical searching time in the following. 

Define a boundary operator $B_{\sigma}: \dC^{E\backslash {M}} \to \dC^{V}$ by 
\[ (B_{\sigma})_{u,e}=
\begin{cases}
1, & u \in e,\\
0, & \text{otherwise}.
\end{cases}
\]
\begin{lem}
Let $P_{M}$ and $B_{\sigma}$ be defined as in the above. Then we have
\begin{equation}
P_{M}=\frac{1}{2(n-1)}(B^{\top}_{\sigma}B_{\sigma}-2I_{|E|-t}). 
\label{PM}
\end{equation}
\label{lem of PM}
\end{lem}
\begin{proof}
Here, $B^{\top}_{\sigma}B_{\sigma}$ is a matrix indexed by $E\backslash{M}$ whose entry is
\begin{align*}
(B^{\top}_{\sigma}B_{\sigma})_{e,f}&=\sum_{u \in V}(B_{\sigma})_{u,e}\cdot (B_{\sigma})_{u,f}\\
&=\sum_{\substack{u \in V\\ u \in e, u \in f}}1
\end{align*}
for $e,f \in E\backslash{M}$. The right-hand-side of the above equation is nothing but the number of the vertex which is an endpoint of both of $e$ and $f$. Thus, we have
\[ (B^{\top}_{\sigma}B_{\sigma})_{e,f}=
\begin{cases}
2, & e=f,\\
1, & e \ne f, \quad e \approx f,\\
0, & \text{otherwise}.
\end{cases}
\] 
Then it holds that $(B^{\top}_{\sigma}B_{\sigma})-2I_{|E|-t}=A(L(G_{M}))$, which completes the proof. 
\end{proof}
In order to estimate the maximum eigenvalue of $P_{M}$, it is useful to obtain that of $B^{\top}_{\sigma}B_{\sigma}$ by Lemma \ref{lem of PM}. As spectra of $B^{\top}_{\sigma}B_{\sigma}$ and $B_{\sigma}B^{\top}_{\sigma}$ are in coincidence except for $0$, we analyze the spectrum of the latter one instead of the former one. 
\begin{lem}
Let $B_{\sigma}$ be defined as in the above. Then it holds that
\[ (B_{\sigma}B^{\top}_{\sigma})_{u, v}=
\begin{cases}
n, & u=v, \quad u \not\in \partial{M},\\
n-1, & u=v, \quad u \in  \partial{M},\\
1, & u \ne v, \quad uv \not\in M,\\
0, & u \ne v, \quad uv \in M.
\end{cases}
\]
\end{lem}
\begin{proof}
It holds that
\begin{align}
(B_{\sigma}B^{\top}_{\sigma})_{u,v}&=\sum_{e \in E\backslash{M}}(B_{\sigma})_{u,e}\cdot (B_{\sigma})_{v,e}\nonumber \\ 
&=\sum_{\substack{e \in E\backslash{M}\\ u \in e, v \in e}}1 
\label{numbers}
\end{align}
for $u, v \in V$. The right-hand-side is the number of edges in $E\backslash{M}$ whose endpoints are $u$ and $v$. We first consider the case of $u=v$. If $u=v \not\in \partial{M}$, then the number of edges in $E\backslash{M}$ whose endpoint is $u$ is $n$ since $u$ is adjacent to all the vertices in $G$. If $u=v \in \partial{M}$, there is only one edge in $M$ whose endpoint is $u$ since $M$ is a matching. Thus, (\ref{numbers}) is $n-1$ in this case. 

We next consider the case of $u \ne v$. If $uv \not\in M$, there is only one edge in $E\backslash{M}$ connecting $u$ and $v$. Thus, (\ref{numbers}) is $1$ in this case. If $uv \in M$, there is no edge in $E\backslash{M}$ whose endpoints are $u$ and $v$, which implies that (\ref{numbers}) is $0$ in this case.
Therefore, we conclude that 
\[ (B_{\sigma}B^{\top}_{\sigma})_{u, v}=
\begin{cases}
n, & u=v, \quad u \not\in \partial{M},\\
n-1, & u=v, \quad u \in \partial{M},\\
1, & u \ne v, \quad uv \not\in M,\\
0, & u \ne v, \quad uv \in M.
\end{cases}
\]
\end{proof}
Thus, by taking a proper labeling of vertices, we express $B_{\sigma}B^{\top}_{\sigma}$ as
\begin{equation}
B_{\sigma}B^{\top}_{\sigma}=\left(
\begin{array}{cc}
A(\mathrm{CP}(t))+(n-1)I_{2t} & J_{2t, n+1-2t}\\
J_{n+1-2t,2t} & A(K_{n+1-2t})+nI_{n+1-2t},
\end{array}
\right),
\label{Bhat}
\end{equation}
where $J_{r,s}$ is the $(r \times s)$-all-one matrix. 
\begin{lem}
Let $B_{\sigma}$ be defined as in the above. Then it holds that 
\[ \mathrm{Spec}(B_{\sigma}B^{\top}_{\sigma})=\{ n-3\}^{t-1}\cup \{ n-1\}^{n-t} \cup \{ \mu_{\pm}\},  \]
where 
\[\mu_{\pm}=\frac{3n-3\pm\sqrt{n^{2}+6n+9-16t}}{2}.\] 
\label{co-random}
\end{lem}
\begin{proof}
Put $\hat{B}=B_{\sigma}B^{\top}_{\sigma}$. As is seen in (\ref{Bhat}), $\hat{B}$ is expressed in terms of the adjacency matrices of $\mathrm{CP}(t)$ and $K_{n+1-2t}$. The spectrum of a complete graph is known to be 
\begin{equation}
\mathrm{Spec}(A(K_{r}))=\{ r-1 \}^{1} \cup \{-1 \}^{r-1},
\label{spec kn}
\end{equation}
for example, see \cite{Spg}. Then we have
\[ \mathrm{Spec}(A(K_{n+1-2t}))=\{ n-2t\}^{1}\cup \{-1 \}^{n-2t}. \]
In addition, it is also known that the eigenvector associated to $(n-2t)$ is ${\bf j}_{n+1-2t}$.  We next analyze the spectrum of $A(\mathrm{CP}(t))$. Since $A(\mathrm{CP}(t))=A(K_{t}) \otimes J_{2,2}$, an eigenvalue of $A(\mathrm{CP}(t))$ is expressed by the product of those of $A(K_{t})$ and $J_{2,2}$. Thus, it follows from $\mathrm{Spec}{(J_{2,2})}=\{0\}^{1}\cup \{2\}^{1}$ and (\ref{spec kn}) that
\[ \mathrm{Spec}(A(\mathrm{CP}(t)))= \{-2 \}^{t-1}\cup \{ 0\}^{t}\cup \{2(t-1)\}^{1}. \]
Similarly, the eigenvector associated to $2(t-1)$ is ${\bf j}_{2t}$. Let $\Psi_{1}=({\bf j}_{2t}, {\bf 0}_{n+1-2t})^{\top} \in \dC^{V}$ and $\Psi_{2}=({\bf 0}_{2t}, {\bf j}_{n+1-2t})^{\top}$. 
Using the above facts, we have 
\begin{align*}
\hat{B}\Psi_{1}&=\left(
\begin{array}{cc}
A(\mathrm{CP}(t))+(n-1)I_{2t} & J_{2t, n+1-2t}\\
J_{n+1-2t,2t} & A(K_{n+1-2t})+nI_{n+1-2t},
\end{array}
\right) \left(
\begin{array}{c}
{\bf j}_{2t}\\
{\bf 0}_{n+1-2t}
\end{array}
\right)\\[3pt]
&=\left(
\begin{array}{c}
(A(\mathrm{CP}(t))+(n-1)I_{2t}){\bf j}_{2t}\\
(J_{n+1-2t,2t}){\bf j}_{2t}
\end{array}
\right)\\[3pt]
&=\left(
\begin{array}{c}
(2(t-1)+n-1){\bf j}_{2t}\\
2t{\bf j}_{n+1-2t}
\end{array}
\right)\\[3pt]
&=(n+2t-3)\Psi_{1}+2t\Psi_{2}
\end{align*}
and
\begin{align*}
\hat{B}\Psi_{2}&=\left(
\begin{array}{cc}
A(\mathrm{CP}(t))+(n-1)I_{2t} & J_{2t, n+1-2t}\\
J_{n+1-2t,2t} & A(K_{n+1-2t})+nI_{n+1-2t},
\end{array}
\right) \left(
\begin{array}{c}
{\bf 0}_{2t}\\
{\bf j}_{n+1-2t}
\end{array}
\right)\\[3pt]
&=\left(
\begin{array}{c}
(J_{2t,n+1-2t}){\bf j}_{n+1-2t}\\
(A(K_{n+1-2t})+nI_{n+1-2t}){\bf j}_{n+1-2t}
\end{array}
\right)\\[3pt]
&=\left(
\begin{array}{c}
(n+1-2t){\bf j}_{2t}\\
(n-2t+n){\bf j}_{n+1-2t}
\end{array}
\right)\\[3pt]
&=(n+1-2t)\Psi_{1}+(2n-2t)\Psi_{2}.
\end{align*}
Thus, $\hat{B}$ is reduced to the following $2 \times 2$-matrix on $\mathrm{Span}\{ \Psi_{1}, \Psi_{2}\}$: 
\begin{equation}
Q=\left(
\begin{array}{cc}
n+2t-3 & n+1-2t\\
2t & 2n-2t
\end{array}
\right).
\label{quotient}
\end{equation}
Here, it holds that $\mathrm{Spec}(Q) \subset \mathrm{Spec}(\hat{B})$. By direct computation, the eigenvalues of $Q$ are  
\[ \mu_{\pm}=\frac{3n-3\pm\sqrt{n^{2}+6n+9-16t}}{2}. \]

Now, we analyze the remaining eigenvalues of $\hat{B}$.  Let $g$ be an eigenvector of $A(K_{n+1-2t})$ associated to $-1$. Then $g$ is orthogonal to ${\bf j}_{n+1-2t}$, see \cite{Spg}. Put $\hat{g}=( {\bf 0}_{2t}, g)^{\top} \in \dC^{V}$. Since $(J_{2t, n+1-2t})g={\bf 0}_{2t}$ and $A(K_{n+1-2t})g=-g$, it is easily checked that $\hat{g}$ is an eigenvector of $\hat{B}$ associated to $n-1$ by similar computation. Taking $g$ as an eigenvector of $A(K_{n+1-2t})$ associated to $-1$, we have thus found $n-2t$ linearly independent eigenvectors of $\hat{B}$ associated to $n-1$.  We next consider the remaining eigenvalues given by those of $A(\mathrm{CP}(t))$. Since $A(\mathrm{CP}(t))$ is a symmetric matrix, eigenvectors associated to $-2$ and $0$ are orthogonal to the one associated to the maximum eigenvalue $2(t-1)$, that is, ${\bf j}_{2t}$.  Let $f$ be an eigenvector of $A(\mathrm{CP}(t))$ associated to $\eta \in \{-2, 0\}$ and $\hat{f}=( f, {\bf 0}_{n+1-2t})^{\top}$. Similarly, it is checked that $\hat{f}$ is an eigenvector of $\hat{B}$ associated to the eigenvalue $\eta+n-1$ for $\eta \in \{-2,0\}$. Recall that the multiplicity of the eigenvalues $-2$ and $0$ of $A(\mathrm{CP}(t))$ are $t-1$ and $t$, respectively. Then the multiplicities of eigenvalues $n-3$ and $n-1$ of $\hat{B}$ are $t-1$ and $t$, respectively.  Therefore, eigenvalues of $\hat{B}$ that we have found are $\mu_{\pm}$, $n-1$ with multiplicity $n-2t+t=n-t$, and $n-3$ with multiplicity $t-1$. 
Since $2+(n-t)+(t-1)=n+1=|V|$, these are all the eigenvalues of $\hat{B}$ and we conclude that 
\[ \mathrm{Spec}(\hat{B})=\{ n-3\}^{t-1}\cup \{ n-1\}^{n-t} \cup \left\{ \frac{3n-3\pm\sqrt{n^{2}+6n+9-16t}}{2}\right\}. \]
Clearly, $n-1 < \frac{3n-3+\sqrt{n^{2}+6n+9-16t}}{2}=\mu_{+}$. Thus, $\mu_{+}$ is the maximum eigenvalue of $\hat{B}$. 
\end{proof}
Note that if $M$ achieves a perfect matching, that is, $t=\frac{n+1}{2}$ for an odd $n$, we have $V\backslash{\partial M}=\phi$ and 
\[ B_{\sigma}B^{\top}_{\sigma}=A(\mathrm{CP}(t))+(n-1)I_{2t}. \]
Then the maximum eigenvalue of $B_{\sigma}B^{\top}_{\sigma}$ is $2(t-1)+(n-1)=2(n-1)$
which coincides with the above $\mu_{+}$ with $t=\frac{n+1}{2}$. Hence, we employ $\mu_{+}$ to the following Theorem for every case. 
\begin{thm}
Let $\alpha, M$ and $P_{M}$ be defined as in the above. Then the classical searching time finding an edge in $M$ is of the order of $n^{2-\alpha}$. 
\end{thm}
\begin{proof}
By Lemma \ref{co-random} and (\ref{PM}), the maximum eigenvalue of $P_{M}$ is 
\begin{align*}
\mu_{m}&=\frac{1}{2(n-1)}(\mu_{+}-2)\\
&=\frac{3n-7+\sqrt{n^{2}+6n+9-16t}}{4(n-1)}
\end{align*}
Since $t=[cn^{\alpha}]=O(n^{\alpha})$ with $0 \le \alpha \le 1$, we have
\[ \sqrt{n^{2}+6n+9-16t}=n+3-8cn^{\alpha-1}+O(n^{\alpha-1}), \]
and
\begin{align*}
\mu_{m}&=\frac{3n-7+\sqrt{n^{2}+6n+9-16t}}{4(n-1)}\\
&=\frac{3n-7+n+3-8cn^{\alpha-1}+O(n^{\alpha-1})}{4(n-1)}\\
&=1-2cn^{\alpha-2}+O(n^{\alpha-2}).
\end{align*}
Therefore, the order of $\frac{1}{1-\mu_{m}}$ is $n^{2-\alpha}$, which completes the proof by \cite{Broder}. 
\end{proof}

\section{Quantum search algorithm}
\subsection{Spectrum of $T_{\sigma}$}
As $G$ is an $n$-regular graph, it follows from (\ref{def of T}) that 
\begin{equation}
(T_{\sigma})_{u,v}=
\begin{cases}
-\frac{1}{n} & \tau(uv)=-1,\\
\frac{1}{n} & \tau(uv)=1,\\
0 & \text{otherwise},
\end{cases}
\label{T-mat}
\end{equation}
which is a {\it weighted adjacency matrix} of a signed complete graph \cite{Akbari}. According to \cite{Akbari}, the spectrum of the adjacency matrix $A$ of a signed complete graph $K_{n+1}$ in which the set of negatively signed edges is a $t$-matching is obtained as follows: 
\begin{description}
\item[(i)] If $t < \lfloor \frac{n+1}{2} \rfloor$, 
\begin{equation}
\mathrm{Spec}(A)=\{-3 \}^{t-1} \cup \{ a_{t}\}^{1} \cup \{-1\}^{n-2t} \cup \{ 1\}^{t} \cup \{b_{t} \}^{1},
\label{adjacency of t-match 1}
\end{equation}
where $a_{t}=\frac{n-3-\sqrt{(n+1)^{2}+4s}}{2}, b_{t}=\frac{n-3+\sqrt{(n+1)^{2}+4s}}{2}$ and $s=n-4t+2$. 
\item[(ii)] If $t=\frac{n+1}{2}$ for an odd $n$,  that is, $M$ achieves a perfect matching, 
 \begin{equation}
\mathrm{Spec}(A)=\{-3 \}^{t-1} \cup \{ 1\}^{t} \cup \{b_{t} \}^{1}.
\label{adjacency of t-match 2}
\end{equation}
Note that $b_{t}=n-2$ in this case.
\end{description} 
Hence, it follows from the fact of $T=\frac{1}{n}A$ that  if $t < \lfloor \frac{n+1}{2} \rfloor$,
\begin{equation}
\mathrm{Spec}(T_{\sigma})=\left\{-\frac{3}{n} \right\}^{t-1} \cup \left\{ \frac{a_{t}}{n} \right\}^{1} \cup \left\{-\frac{1}{n} \right\}^{n-2t} \cup \left\{ \frac{1}{n} \right\}^{t} \cup \left\{ \frac{b_{t}}{n} \right\}^{1}.
\label{transition of t-match 1}
\end{equation}
Moreover, the normalized eigenvector of $T_{\sigma}$ associated to the maximum eigenvalue $\lambda_{m}=\frac{b_{t}}{n}$ is
\begin{equation}
f_{M}=\frac{1}{\sqrt{c_{n}}}(\underbrace{\rho_{n}, \dots, \rho_{n}}_{2t}, \underbrace{1, \dots, 1}_{n+1-2t} )^{\top}, 
\label{eig of T}
\end{equation}
where $\rho_{n}=\frac{-(s+1)+\sqrt{\Delta_{n}}}{4t}, \Delta_{n}=(n+1)^{2}+4s$ and $c_{n}=2t\rho^{2}_{n}+n+1-2t$ by \cite{Akbari}. 
If $t=\frac{n+1}{2}$ for an odd $n$, it holds
\begin{equation}
\mathrm{Spec}(T_{\sigma})=\left\{-\frac{3}{n} \right\}^{t-1} \cup \left\{ \frac{1}{n} \right\}^{t} \cup \left\{\frac{b_{t}}{n} \right\}^{1}
\label{transition of t-match 2}
\end{equation} 
and the normalized eigenvector of $T_{\sigma}$ associated to $\lambda_{m}=\frac{b_{t}}{n}$ is $\frac{1}{\sqrt{n+1}}{\bf j}_{n+1}$ by \cite{Akbari}. Remark that $\rho_{n}=1, \sqrt{\Delta_{n}}=(n-1)^{2}$ and $c_{n}=n+1$ in this case.


\subsection{Proof of Theorem \ref{search on Kn+1}}
First, we consider the case where $M$ does not achieve a perfect matching. Let us estimate the parameters  $\sqrt{\Delta_{n}}=\sqrt{(n+1)^{2}+4s}$, $\rho_{n}=\frac{-(s+1)+\sqrt{\Delta_{n}}}{4t}$ and $c_{n}=2t\rho^{2}+n+1-2t$. 
First, $\sqrt{\Delta_{n}}$ is expanded as
\[ \sqrt{\Delta_{n}}=n+3-8cn^{\alpha-1}+O(n^{\alpha-2}). \]
Hence, we have
\begin{align*}
\rho_{n}&=\frac{-(s+1)+\sqrt{\Delta_{n}}}{4t}\\
&=\frac{-(n-4t+3)+(n+3-8cn^{\alpha-1}+O(n^{\alpha-2}))}{4t}\\
&=1+O\left(\frac{1}{n}\right)
\end{align*}
and 
\begin{align*}
c_{n}&=2t\rho^{2}_{n}+n+1-2t\\
&=n+1+2t (\rho^{2}_{n}-1).\\
&=n+O(1).
\end{align*}  
Then, for a sufficiently large $n$, the values $\rho_{n}$ and $c_{n}$ are approximated as $1$ and $n$, respectively. 

Next, let us analyze the eigenvector of $U_{\sigma}$ associated to the eigenvalue induced by $\lambda_{m}$. To this end, we decompose $\mathcal{A}$ into six distinct subsets $\mathcal{A}_{1}, \mathcal{A}_{2}, \mathcal{A}_{3}, \mathcal{A}_{4}, \mathcal{A}_{5}, \mathcal{A}_{6}$ as
\begin{align*}
&\mathcal{A}_{1}=\mathcal{M},\\
&\mathcal{A}_{2}=\mathcal{M}^{-1},\\
&\mathcal{A}_{3}=\{ a \in \mathcal{A} \mid a \not\in \mathcal{M}\cup \mathcal{M}^{-1}, \quad t(a), o(a) \in \partial{M}\},\\
&\mathcal{A}_{4}=\{a \in \mathcal{A} \mid t(a) \in \partial{M}, o(a) \not\in \partial{M}\},\\
&\mathcal{A}_{5}=\{a \in \mathcal{A} \mid t(a) \not\in \partial{M}, o(a) \in \partial{M}\},\\
&\mathcal{A}_{6}=\{a \in \mathcal{A} \mid t(a) \not\in \partial{M}, o(a) \not\in \partial{M}\}.
\end{align*}
Put $\theta_{m}=\cos^{-1}{\lambda_{m}}$.  Inserting (\ref{eig of T}) into (\ref{eig func}), we obtain the normalized eigenvector $\varphi_{\pm \theta_{m}}$ of $U_{\sigma}$ associated to $e^{\pm i \theta_{m}}$ as follows: 
\begin{equation}
\varphi_{\pm \theta_{m}}(a)=\frac{1}{\sqrt{2nc_{n}}\sin{\theta_{m}}}\times
\begin{cases}
-\rho_{n}(1+e^{\pm i \theta_{m}}), & a \in \mathcal{A}_{1}, \\[3pt]
\rho_{n}(1+e^{\pm i \theta_{m}}), & a \in \mathcal{A}_{2}, \\[3pt]
\rho_{n}(1-e^{\pm i \theta_{m}}), & a \in \mathcal{A}_{3}, \\[3pt]
\rho_{n}-e^{\pm i \theta_{m}}, & a \in \mathcal{A}_{4}, \\[3pt]
1-\rho_{n}e^{\pm i \theta_{m}}, & a \in \mathcal{A}_{5}, \\[3pt]
1-e^{\pm i \theta_{m}}, & a \in \mathcal{A}_{6}.
\end{cases}
\label{varphi}
\end{equation}
Here, we will count the number of arcs satisfying each condition as in the above. Clearly, both of the numbers of arcs in $\mathcal{A}_{1}$ and $\mathcal{A}_{2}$ are $t$. For a pair of two distinct arcs $(u,v), (w,x)$ in $\mathcal{M}$, the arcs $(u, w), (u, x), (v, w), (v, x)$ and their inverse arcs satisfy the condition as in $\mathcal{A}_{3}$. Thus, the number of arcs in $\mathcal{A}_{3}$ is $8 \times \binom{t}{2}=4t(t-1)$. For an arc $a$ in $\mathcal{M}$, the number of arcs $b (\ne a)$ satisfying $t(b)=t(a)$ and $o(b) \not\in \partial{M}$ is same as the number of vertices in $V\backslash{\partial{M}}$, that is, $n+1-2t$. Similarly, the number of arcs $b (\ne a)$ satisfying $t(b)=o(a)$ and $o(b) \not\in \partial{M}$ is $n+1-2t$. Thus, the number of arcs in $\mathcal{A}_{4}$ is $(n+1-2t) \times 2 \times t=2t(n+1-2t)$, which is same as $|\mathcal{A}_{5}|$. Finally, the number of the other arcs is $|\mathcal{A}|-\sum^{5}_{i=1}|\mathcal{A}_{i}|=n(n+1)-2t-4t(t-1)-4t(n+1-2t)=n^{2}+n+4t^{2}-4nt-2t$. 
We make a list of the numbers of arcs satisfying the above conditions as follows:
\begin{table}[htbp]
\begin{tabular}{|c|c|c|c|c|c|}\hline
$\mathcal{A}_{1}$ & $\mathcal{A}_{2}$ & $\mathcal{A}_{3}$ & $\mathcal{A}_{4}$ & $\mathcal{A}_{5}$ & $\mathcal{A}_{6}$\\ \hline
$t$ & $t$ & $4t(t-1)$ & $2t(n+1-2t)$ & $2t(n+1-2t)$ & $n^{2}+n+4t^{2}-4nt-2t$\\ \hline
\end{tabular}
\end{table}

Let $u \in \dC^{\mathcal{A}}$ be the uniform state, that is,
\begin{eqnarray*}
u(a)=\sqrt{\frac{1}{n(n+1)}}, & a \in \mathcal{A}(G).
\end{eqnarray*}
Now, we employ $u$ as the initial state. In this paper, we estimate the total of the time complexity of our quantum searching model based on \cite{Portugal}. In other words, we obtain the order of the time  at which the finding probability on the marked edges is sufficiently high. We give the outline of the proof as follows:
\begin{itemize}
\item[(1)] We define two vectors $\beta_{+}$ and $\beta_{-}$ and see that $\beta_{-}$ is sufficiently close to $u$. 
\item[(2)] We estimate the order of the time complexity $k_{f}$ such that $U^{k_{f}}_{\sigma}\beta_{-}$ is close to $\beta_{+}$. 
\item[(3)] We estimate the order of the finding probability $FP_{n}$ on the marked edges of $\beta_{+}$. 
\end{itemize}
It is known that the total of the time complexity is obtained as 
\[  k_{f} \times \sqrt{\frac{1}{FP_{n}}} \]
by the {\it amplitude amplification} \cite{Brassard}. Multiplying the orders of $k_{f}$ and $FP_{n}$, we get the order of the total of the time complexity. In order to complete the proof, we get their orders.

Define 
\[ \beta_{\pm}:=\frac{1}{\sqrt{2}}\left(\varphi_{+\theta_{m}} \pm \varphi_{-\theta_{m}} \right). \]
Then we have
\begin{equation*}
\beta_{+}(a)=\frac{1}{\sqrt{nc_{n}}\sin{\theta_{m}}}\times
\begin{cases}
-\rho_{n}(1+\cos{\theta_{m}}), & a \in \mathcal{A}_{1},\\
\rho_{n}(1+\cos{\theta_{m}}), & a \in \mathcal{A}_{2},\\
\rho_{n}(1-\cos{\theta_{m}}), & a \in \mathcal{A}_{3},\\
\rho_{n}-\cos{\theta_{m}}, & a \in \mathcal{A}_{4},\\
1-\rho_{n}\cos{\theta_{m}}, & a \in \mathcal{A}_{5},\\
1-\cos{\theta_{m}}, & a \in \mathcal{A}_{6},
\end{cases}
\end{equation*}
and
\begin{equation*}
\beta_{-}(a)=\frac{1}{\sqrt{nc_{n}}}\times
\begin{cases}
-i\rho_{n}, & a \in \mathcal{A}_{1} \cup \mathcal{A}_{3} \cup \mathcal{A}_{5}, \\[3pt]
i\rho_{n}, & a \in \mathcal{A}_{2}, \\[3pt]
-i, & a \in \mathcal{A}_{4} \cup \mathcal{A}_{6},
\end{cases}
\end{equation*}
by (\ref{varphi}).  Let us analyze the overlap between $\beta_{-}$ and $u$. It holds that 
\begin{align*}
|\langle u, \beta_{-} \rangle|&=\left| \frac{1}{n\sqrt{c_{n}(n+1)}}\{(|\mathcal{A}_{1}|+|\mathcal{A}_{3}|+|\mathcal{A}_{5}|)\times(-i\rho_{n})+|\mathcal{A}_{2}| \times (i\rho_{n})+(|\mathcal{A}_{4}|+|\mathcal{
A}_{6}|)\times (-i) \} \right| \nonumber \\[3pt] 
&=\left| \frac{1}{n\sqrt{c_{n}(n+1)}}\{(2nt-t)\times(-i\rho_{n})+t \times (i\rho_{n})+n(n-2t+1)\times (-i) \} \right| \nonumber \\[3pt] 
&=\left| \frac{1}{n\sqrt{c_{n}(n+1)}} \{(2t-2nt)\rho_{n}-(n^{2}-2nt+n) \} \right|. 
\end{align*}
As $\rho_{n} \approx 1$ and $c_{n} \approx n$, we have 
\[ |\langle u, \beta_{-} \rangle| = 1-O\left(\frac{1}{n}\right). \]
Thus, $\beta_{-}$ is so close to $u$ and we regard $\beta_{-}$  as the initial state.  Let
\[ \psi_{k}=U^{k}_{\sigma}\beta_{-}=\frac{1}{\sqrt{2}}\left( e^{i \theta_{m}k}\varphi_{+\theta_{m}}-e^{-i \theta_{m}k}\varphi_{-\theta_{m}}\right) \]
and $k_{f}=\lfloor \frac{\pi}{2\theta_{m}} \rfloor$. Then  $\psi_{k_{f}}$ is regarded as $i\beta_{+}$ since $e^{\pm i \theta_{m}k_{f}}$ is close to $\pm i$. Now, we compute the time complexity converting $\beta_{-}$ to its orthogonal vector $i\beta_{+}$, that is, the order of $\frac{1}{\theta_{m}}$. To this end, let us estimate $\theta_{m}$ for a sufficiently large $n$.  Since $\lambda_{m}=\cos{\theta_{m}}$ tends to $1$ as $n \to \infty$, $\theta_{m}$ is approximated as $\sin{\theta_{m}}$. Then we have
\begin{align*}
\theta_{m} &\approx \sin{\theta_{m}}=\sqrt{1-\cos^{2}{\theta_{m}}}\\[3pt]
&=\sqrt{1-\left(\frac{n-3+\sqrt{\Delta_{n}}}{2n}\right)^{2}}\\[3pt]
&=\sqrt{1-\left(\frac{2n-8cn^{\alpha-1}+O(n^{\alpha-2})}{2n}\right)^{2}}\\[3pt]
&=\sqrt{8c}n^{\frac{\alpha-2}{2}}+o\left(n^{\frac{\alpha-2}{2}}\right). 
\end{align*}
Then the order of $k_{f}$ is $n^{\frac{2-\alpha}{2}}$.  

At this time, the finding probabilities on arcs $a \in \mathcal{M}$ and $a^{-1} \in \mathcal{M}^{-1}$ are close to 
\begin{equation} 
|\beta_{+}(a)|^{2}=|\beta_{+}(a^{-1})|^{2}=\left| \frac{\rho_{n}}{\sqrt{nc_{n}}\sin{\theta_{m}}}(1+\cos{\theta_{m}}) \right|^{2}
\label{FP}
\end{equation}
by (\ref{varphi}). As $c_{n} \approx n, \rho_{n} \approx 1, \cos{\theta_{m}} \approx 1$ and  $\sin{\theta_{m}} \approx \sqrt{8c}n^{\frac{\alpha-2}{2}}$, (\ref{FP}) is of the order of $n^{-\alpha}$. Hence, the finding probability on the edges in $M$ is
\begin{align*}
FP_{n}&= \sum_{a \in \mathcal{M}}\left( |\beta_{+}(a)|^{2}+|\beta_{+}(a^{-1})|^{2}\right)\\
&=2t \cdot \left| \frac{\rho_{n}}{\sqrt{nc_{n}}\sin{\theta_{m}}}(1+\cos{\theta_{m}}) \right|^{2}\\[3pt]
&\approx 2cn^{\alpha} \cdot \left| \frac{1}{\sqrt{n\cdot n}\sqrt{8c}n^{\frac{\alpha-2}{2}}}(1+1)\right|^{2}\\[3pt]
&=1-o(1). 
\end{align*}
By the amplitude amplification, the total of the time complexity $k_{total}$ is given by
\[ k_{total} = k_{f} \times \sqrt{\frac{1}{FP_{n}}}. \]
Recall that the orders of $k_{f}$ and $FP_{n}$ are $n^{\frac{2-\alpha}{2}}$ and $1$, respectively. Therefore, the order  of  $k_{total}$ becomes $n^{\frac{2-\alpha}{2}}$. 

In the case where $M$ achieves a perfect matching, it is similarly shown that the total of the time complexity is $O(n^{\frac{1}{2}})$ by repeating the same argument with $\rho_{n}=1, c_{n}=n+1$ and  $\sqrt{\Delta_{n}}=(n-1)^{2}$. Therefore, we complete the proof.

\section{Summary and discussion}
In this paper, we introduced a signed graph whose edges are negatively signed. Especially, we proposed a quantum searching algorithm in a signed complete graph $K_{n+1}$ and it enable us to find negatively signed edges, say, marked edges, quadratically faster than a classical algorithm, where the number of marked edges now is $t=O(n^{\alpha})$ for $0 \le \alpha \le 1$ and the set of marked ones becomes a $t$-matching which is a combination of disjoint $t$ edges. In other words, we found a quantum search algorithm detecting a marked edge within $O(n^{\frac{2-\alpha}{2}})$ while an algorithm based on a classical random walk requires the time complexity of $O(n^{2-\alpha})$. 
The algorithm proposed in this paper only reveals  quadratic speed-up to find one marked edge from a matching. Hence, considering an algorithm to know how marked edges are located is one of our future problem.  An extension of our proposed algorithm to the other graphs is also still open. 

\section*{Acknowledgement}
E.S. acknowledges financial supports from Japan Society for the Promotion
of Science Grant-in-Aid for Scientific Research (C) 19K03616, and Research Origin for
Dressed Photon.

\end{document}